\documentclass[useAMS,usenatbib]{mn2e}
\usepackage{graphicx}
\usepackage{latexsym}

\title{Ionisation-induced star formation I: The collect and collapse model}

\author[J. E. Dale, I. A. Bonnell, A. P. Whitworth]{J. E. Dale$^{1}$\thanks{E-mail: Jim.Dale@astro.le.ac.uk (JED)},
I. A. Bonnell$^{2}$A. P. Whitworth$^{3}$\\
$^{1}$Department of Physics and Astronomy, University of Leicester, University Road, Leicester, LE1 7RH\\
$^{2}$Department of Physics and Astronomy, University of St Andrews, North Haugh, St Andrews, Fife KY16 9SS\\
$^{3}$Department of Physics and Astronomy, Cardiff University, Cardiff, CF24 3AA}

\begin{document}

\pagerange{\pageref{firstpage}--\pageref{lastpage}} \pubyear{2006}

\maketitle

\label{firstpage}

\def\mnras{MNRAS}
\def\apj{ApJ}
\def\aap{A\&A}
\def\apjl{ApJL}
\def\apjs{ApJS}
\def\bain{BAIN}

\begin{abstract}
We conduct Smoothed Particle Hydrodynamics simulations of the `collect and collapse' scenario \citep{1977ApJ...214..725E} for star formation triggered by an expanding HII region. We simulate the evolution of a spherical uniform molecular cloud with an ionising source at its centre. The gas in the cloud is self-gravitating, although the cloud is prevented from globally collapsing. We find that the shell driven by the HII region fragments to form numerous self--gravitating objects. We repeat our calculations at four numerical resolutions to ensure that they are converged. We compare our results to the analytical model of \cite{1994MNRAS.268..291W} and show that our simulations and the predictions of Whitworth et al are in good agreement in the sense that the shell fragments at the time and radius predicted by \cite{1994MNRAS.268..291W} to within $20\%$ and $25\%$ respectively. Most of the fragments produced in our two highest resolution calculations are approximately half the mass of those predicted by \cite{1994MNRAS.268..291W}, but this conclusion is robust against both numerical resolution and the presence of random noise (local fluctuations in density of a factor of $\sim2$) in the initial gas distribution. We conclude that such noise has little impact on the fragmentation process.
\end{abstract}

\begin{keywords}
stars: formation, ISM: HII regions
\end{keywords}
 
\section{Introduction}
The question of whether the extreme mechanical and radiative luminosities of OB stars can trigger or induce the birth of other stars is a crucial and still open question in the field of star formation. The winds, expanding HII regions and eventual supernova explosions generated by massive stars sweep up and compress the surrounding gas in their natal molecular clouds. The Jeans mass, M$_{J}$, defining the maximum mass of gas at a given temperature stable against gravitational collapse, is related to the gas density $\rho$ as M$_{J}\propto 1/\sqrt{\rho}$ (for a spherical object). Isothermal compression of molecular gas therefore decreases the Jeans mass, encouraging gravitational fragmentation and, ultimately, star formation.\\
\indent The `collect and collapse' model, first proposed by \cite{1977ApJ...214..725E}, is one of the simplest models for induced star formation, since it is essentially one--dimensional. It therefore provides a useful test--case in which theoretical, numerical and observational analyses may be relatively easily compared.\\
\indent When an ionising source is suddenly turned on within a uniform gas distribution, the initially--plentiful supply of photons drives a spherical ionisation front radially outward from the source at a velocity much higher than the speed of sound in the cold neutral gas. This is known as the R-type phase of the front's evolution. (\cite{1954BAN....12..187K}, \cite{1974agn..book.....O}) As the volume of gas behind the ionisation front grows, recombinations consume a progressively larger fraction of the ionising photons and the propagation speed of the front drops. When the expansion speed of the front becomes comparable to the sound speed in the gas, the front transitions from being R-type to D-type in which the bubble of ionised gas expands at the sound speed in the \textit{ionised} gas ($\sim10$ kms$^{-1}$ for hydrogen at $10^{4}$ K). The radius at which this transition occurs is the Str\"omgren radius, R$_{s}$, given by
\begin{eqnarray}
R_{s}=\sqrt[3]{\frac{L_{*}}{4\pi \alpha_{B} n^{2}}}\sim0.65\left(\frac{L{*}}{10^{49}\textrm{s}^{-1}}\right)^{1/3}\left(\frac{n}{10^{3}\textrm{cm}^{-3}}\right)^{-2/3}\textrm{pc},
\end{eqnarray}
where $L_{*}$ is the source luminosity in ionising photons per second, $n$ is the initial number density and $\alpha_{B}$ is a recombination coefficient, defined such that the recombination rate per unit volume to all atomic states except the ground state is $\alpha_{B} n^{2}$. Recombinations direct to the ground state, which produce secondary ionising photons, are treated by the `on--the--spot' approximation -- secondary ionising photons are assumed to be absorbed immediately very close to their site of emission. This assumption is achieved by modifying the true recombination coefficient $\alpha$, yielding the modified recombination coefficient $\alpha_{B}$. The Str\"omgren radius thus simply defines the volume in which all the ionising photons emitted by the source are consumed by recombinations within the ionised gas, assuming that the gas density is unchanged by expansion.\\
\indent The expansion of the HII region through the surrounding neutral gas is highly supersonic and drives a shock, sweeping up an increasingly massive and dense shell of cool neutral gas - this is the `collect' phase of the `collect--and--collapse' process in which the HII region simply acts like a snowplough.\\
\indent If the expansion of the HII region continues for long enough, the surface density of the shell increases to the point where the shell becomes self--gravitating. The shell is then expected to fragment and individual fragments may then enter a non--linear collapse phase, possibly forming stars. This feature of HII region expansion has been rather less well studied, owing to the non-linear nature of the evolution at this stage. \cite{1994MNRAS.268..291W} (hereafter usually referred to as W94) studied the evolution of an infinite uniform volume of gas in which an ionising radiation source is suddenly ignited. They used a two--dimensional Jeans analysis to study the evolution of the dense shell swept up by the HII region. In the next section, we describe W94's model, since this forms the basis of our simulations. In Section 3, we describe our numerical methods and simulations. We present our results in Section 4 and draw our conclusions in Section 5.\\

\section{A theoretical collect--and--collapse model}
\cite{1994MNRAS.268..291W} consider the gravitational stability of the shocked shell driven by an HII region expanding in a uniform medium of density $\rho_{0}$. We summarise their methodology here - for a more detailed exposition, see their Section 5 and Appendix B.\\
\indent W94 represent the time evolution of the radius of the HII region, and therefore of the shocked shell, as a power law given by $R(t)=Kt^{\beta}$. In a uniform medium, $K\approx(7a_{HII}/4)^{4/7}R_{s}^{-1}$ and $\beta=4/7$ (e.g. \cite{1978ppim.book.....S}). For $R(t)>$a few$\times R_{s}$, nearly all the material with $r>R(t)$ is neutral material in the swept--up shell, so the surface density $\Sigma$ of the shell is $\approx\rho_{0}R(t)/3$. They consider a small circular element of the shell and estimate the inward (i.e. azimuthal) acceleration experienced by the fragment, which is a function of the fragment's self--gravity and internal pressure, and of the velocity divergence due to the expansion of the shell. They derive the shortest timescale, $t_{frag}$ on which such an element will begin to condense out of the shell and take fragmentation to begin at $t=t_{frag}$. Using the power--law expression for the expansion of  the shell, they derive expressions for the radius of the shell when fragmentation begins $R_{frag}$, and the mass $M_{frag}$ of these fragments. For convenience, we reproduce the expressions derived by W94 below. In these expressions, $a_{.2}$ is the sound speed inside the shocked layer in units of $0.2$ km s$^{-1}$, $L_{49}$ is the source ionising flux in units of $10^{49}$ photons s$^{-1}$, and $n_{3}$ is the initial gas \textit{atomic} number density in units of $10^{3}$ cm$^{-3}$. In these simulations, we consider gas which is initially pure molecular hydrogen, so the effective atomic number density is simply twice the molecular number density.\\
\begin{eqnarray}
R_{frag}\sim5.8a_{.2}^{4/11}L_{49}^{1/11}n_{3}^{-6/11}\textrm{pc}
\label{eqn:rfrag}
\end{eqnarray}
\begin{eqnarray}
t_{frag}\sim1.6a_{.2}^{7/11}L_{49}^{-1/11}n_{3}^{-5/11}\textrm{Myr}
\label{eqn:tfrag}
\end{eqnarray}
\begin{eqnarray}
M_{frag}\sim23a_{.2}^{40/11}L_{49}^{-1/11}n_{3}^{-5/11}\textrm{M}_{\odot}
\label{eqn:mfrag}
\end{eqnarray}

\section{Numerical methods and simulations}
\begin{figure}
\includegraphics[width=0.50\textwidth]{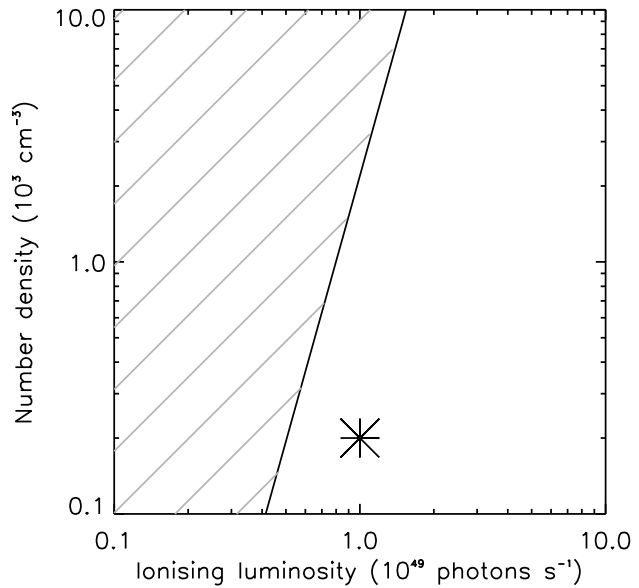}
\caption{Plot of the luminosity--number density parameter space with the constraint that the initial Str\"omgren radius be at least one--tenth the fragmentation radius. The hatched area is forbidden by this constraint and the asterisk marks the parameters used in our calculations.}
\label{fig:param_space}
\end{figure}

We conduct our simulations using the Smoothed Particle Hydrodynamics code described in detail by \cite{1995MNRAS.277..362B}. This code is a hybrid N--body--SPH code in which dense, gravitationally--bound and contracting clumps of gas particles can be replaced by point masses called sink particles. To increase efficiency, particles are integrated on their own individual timesteps. We modified the code to allow the inclusion of point sources of ionising radiation, which may be sink particles or arbitrarily--chosen points in the simulation volume. Our algorithm defines a `Str\"omgren volume' around the ionising source and heats the gas within that volume to $10^{4}$ K. The particles are first sorted by increasing radius from the source. Each particle is then assessed to see whether it lies in front of or behind the ionisation front. The $i$ particles nearest to the vector between the `target' particle and the source are located using a technique similar to that presented in \cite{2000MNRAS.315..713K}. The positions and densities of these $i$ particles are projected onto this vector so that the density $\rho_{i}$ is known at the positions $r_{i}$ along the vector. The ionising photon flux per unit solid angle $S$ reaching the target particle is then calculated as
\begin{eqnarray}
S=\frac{L_{*}}{4\pi}-\sum_{i}\alpha_{B}\frac{(\rho_{i}+\rho_{(i-1)})}{2}r_{(i-1)}^{2}(r_{i}-r_{(i-1)}).
\end{eqnarray}
The zeroth particle in the sum is the ionising source, whose density is taken to be zero. This approximation is equivalent to taking the density variations along this vector to represent the radial variations in a spherically-symmetric gas distribution, and taking each line--of--sight to be independent, consistent with the on--the--spot approximation. If $S>0$, so that a non--zero ionising flux is reaching the target particle, the particle is ionised and heated. A value of $S$ less than zero implies that all the ionising photons emitted in the direction of the target particle are absorbed before reaching it, so that the particle is beyond the ionisation front.\\
\indent In the simulations presented here, the Str\"omgren volume is updated every $270$ yr (i.e. $\sim10^{4}$ times per simulation). Increasing or decreasing the ionisation timestep by factors of two has no discernible effect on either the results or the runtimes of our simulations, so the choice of timestep is arbitrary. There is presumably a maximum ionisation timestep which, if exceeded, no longer gives consistent results, but we did not investigate this.\\
\indent Since simulation of an infinite volume of gas is impossible, and periodic boundary conditions are inappropriate for this problem, we had to construct a finite volume of gas in which all the physical phenomena in which we are interested could be contained and adequately numerically resolved, while still making use of realistic quantities for variables of astrophysical importance, e.g. gas density and source luminosity.\\
\indent We used a finite spherical gas distribution, pressure--confined at its outer boundary. The gas is pure molecular hydrogen initially. We prevented the cloud from radially collapsing by exerting a fictitious force on each particle equal in magnitude but opposite in sign to the gravitational force on the particle due to all the material interior to the particle. We evolved a test cloud for a freefall time without an ionising source to confirm that the initial conditions were stable and found no detectable particle movement save for a thin layer on the edge of the cloud where particles moved slightly from their initial positions. The resulting change in density over a freefall time was very small and not important, since we intended to terminate our simulations before the shock reached the edge of the cloud. The fictitious force was also present during our fragmentation calculations. The presence of the force will not affect the fragmentation process itself, since the expanding shell is confined by thermal pressure from the HII region and ram pressure from the undisturbed gas, not by self--gravity for the duration of our simulations. The fragmentation process is then essentially two--dimensional, as self--gravity is effective only in tangential directions.\\
\indent The characteristics of the fragmentation process depend on three parameters - the isothermal sound speed in the neutral gas, the source ionising luminosity and the initial gas number density. This results in a large parameter space, some of which is not accessible to study numerically due either to numerical resolution issues or prohibitively long calculation times. We therefore do not intend to explore this space thoroughly in this paper. Instead, we select a point in the parameter space for which we may perform calculations over a range of numerical resolutions in a reasonable timeframe and compare our results with the theoretical predictions of \cite{1994MNRAS.268..291W}. In order to reduce the parameter space from three dimensions to two, we choose to simulate molecular clouds consisting of pure molecular hydrogen at a temperature of $10$K. This fixes the isothermal sound speed at $0.2$ km s$^{-1}$, so that $a_{.2}=$ unity. We now plot the two dimensional parameter space and shade regions which satisfy two numerically--driven constraints. The first constraint is that the initial Str\"omgren radius be no less than one tenth of the fragmentation radius, ensuring adequate resolution of the HII region. The second constraint is that the ratio of the fragment mass to the cloud mass must exceed $100/N_{part}$, where $N_{part}$ is the number of particles in the simulation. This constraint ensures that the fragments are adequately resolved by at least one hundred particles, a limit recommended by \cite{1997MNRAS.288.1060B}. \cite{1997MNRAS.288.1060B} stress that, to resolve gravitational fragmentation correctly in SPH, the Jeans mass at which fragmentation occurs should be equal to the mass of at least twice the mean number of SPH particle neighbours, i.e. $100$ particles. \cite{2006A&A...450..881H} find that even gross violation of this condition does not lead to spurious fragmentation, although the collapse of marginally--unstable (i.e. slightly larger than the Jeans mass) objects is suppressed. Based on the predictions of W94, we selected initial conditions and numerical resolutions so that \cite{1997MNRAS.288.1060B}'s limit would be comfortably met.\\
\indent We can express $R_{s}/R_{frag}$ and $M_{frag}/M_{total}$ as \\
\begin{eqnarray}
\frac{R_{s}}{R_{frag}}\sim8.0a_{.2}^{4/11}L_{49}^{10/33}n_{3}^{-28/11}
\label{eqn:r_res}
\end{eqnarray}
\begin{eqnarray}
\frac{M_{frag}}{M_{total}}\sim0.22a_{.2}^{28/11}L_{49}^{-4/11}n_{3}^{-24/11}
\label{eqn:m_res}
\end{eqnarray}
\indent The fragment mass is a fairly weak constraint in a simulation where several $\times10^{5}$ particles can be used, so we ignore it here. We plot the parameter space with the constraint on $R_{frag}$ in Figure \ref{fig:param_space}.\\
\indent We choose a source luminosity of $10^{49}$ s$^{-1}$, consistent with an O7--type star, and an initial neutral gas molecular density of $100$ cm$^{-3}$ (corresponding to an atomic number density in the HII region of $200$ cm$^{-3}$), representative of low--density molecular clouds. Making use of the analysis by \cite{1994MNRAS.268..291W}, we then obtain the radius at which the shocked shell should fragment as $R_{frag}\simeq14$ pc, and an expected fragment mass of $M_{frag}\simeq48$ M$_{\odot}$. For convenience, we record the physical parameters of the cloud and radiation source and the expected characteristics of the fragmentation process in Table \ref{table:init_params}.\\
\indent In numerical work it is desirable to repeat simulations over a range of numerical resolutions (a range of particle numbers in the case of SPH) to ensure that convergence is achieved and that the physical processes of interest are being resolved. We therefore conducted a series of simulations using the physical variables given in Table \ref{table:init_params} with particle numbers of $0.37, 0.74, 1.3$ and $2.6\times 10^{6}$ particles. Note that the smallest mass that can be resolved in our lowest--resolution calculation is $\sim17$ M$_{\odot}$, well below the expected fragment mass. The $2.6$ million--particle simulation took about six weeks to run on a $4\times2.4$ GHz SUN v40z machine.\\
\indent\\
\begin{table}
\begin{tabular}{|l|l|}
\hline
$L_{*}$ & $10^{49}$ s$^{-1}$\\
\hline
$n$ & $200$ cm$^{-3}$\\
\hline
$R_{cloud}$ & $14.6$ pc\\
\hline
$M_{cloud}$ & $6.4\times10^{4}$ M$_{\odot}$\\
\hline
$R_{frag}$ & $13.9$ pc\\
\hline
$t_{frag}$ & $3.3\times10^{6}$ yr\\
\hline
$M_{frag}$ & $48$ M$_{\odot}$\\
\hline
\end{tabular}
\caption{The physical parameters of the cloud and ionising source used in our calculations, and the predictions for the radius and time at which the shocked shell should fragment, and the mass of the fragments, derived from Equations 2--4.}
\label{table:init_params}
\end{table}

\section{Results}
\subsection{Numerical convergence}
\begin{figure}
\includegraphics[width=0.5\textwidth]{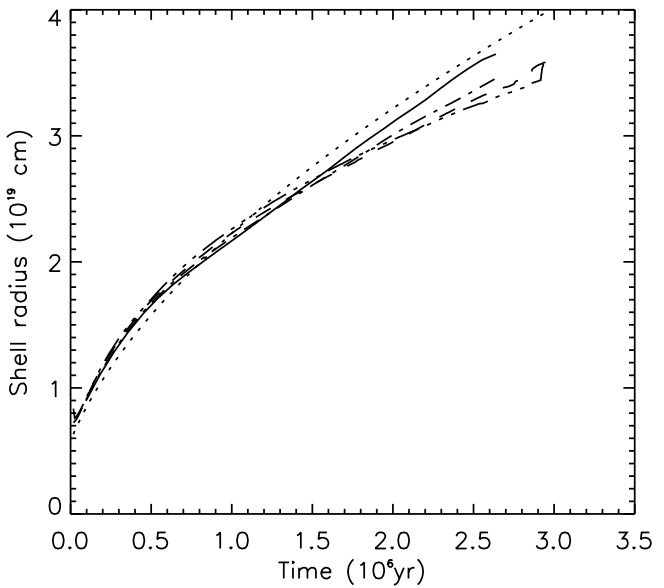}
\caption{Comparison of the evolution of the radius of the shocked shell in our four calculations. Lines correspond to particle numbers (in units of $10^{6}$ particles) as follows: dot--dot--dot--dashed line: 0.37, dashed line: 0.74, dot--dashed line: 1.3, solid line: 2.6. We also plot the Spitzer solution for this system (dotted line)}
\label{fig:hii_compare}
\end{figure}

\begin{figure}
\includegraphics[width=0.5\textwidth]{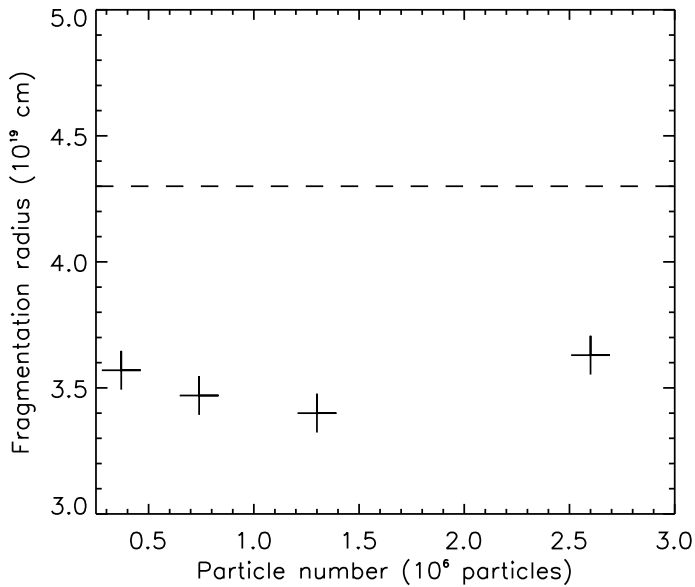}
\caption{Comparison of the radius at which the first point mass is formed in the four simulations (crosses) with the prediction made by W94 (dashed line).}
\label{fig:rfrag}
\end{figure}

\begin{figure}
\includegraphics[width=0.5\textwidth]{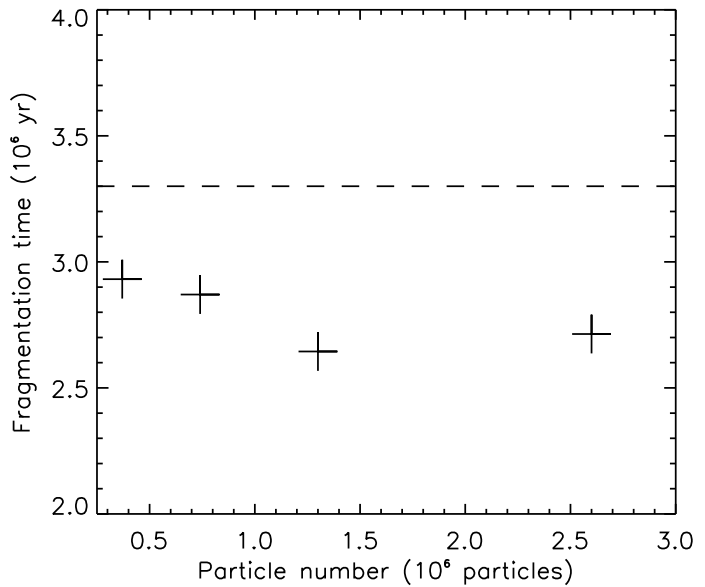}
\caption{Comparison of the time at which the first point mass is formed in the four simulations (crosses) with the prediction made by W94 (dashed line).}
\label{fig:tfrag}
\end{figure}

In its early stages, the evolution of the system under consideration here is essentially one dimensional, so we first examine the growth of the HII region in each of our calculations and compare it with the well--known Spitzer solution for the evolution of the HII region radius R$(t)$ \citep{1978ppim.book.....S}:
\begin{eqnarray}
R(t)=R_{s}\left(1+\frac{7}{4}\frac{c_{s}t}{R_{s}}\right)^{\frac{4}{7}},
\end{eqnarray}
where $c_{s}$ is the sound speed in the ionised gas.\\
\indent In Figure \ref{fig:hii_compare} we plot the evolution of the radius of the shocked shell with time in our four calculations. The radius of the shell is estimated by taking the average distance of the $10^{3}$ densest particles in each simulation from the radiation source at $r=0$, since these particles should reliably reveal the location of the shocked shell outside the ionisation front.\\
\indent The agreement of the four runs with each other is impressive in the early stages of the simulations. The four plots are initially very difficult to distinguish, although they do diverge slightly as the time at which fragmentation occurs ($\sim3\times10^{6}$ yr) is approached. We also note that the agreement between our calculations and the analytic solution improves somewhat as the particle number increases. The evolution of the shells clearly agrees well with the analytical solution of \cite{1978ppim.book.....S}.\\
\indent The fragmentation process itself is rather more difficult to analyse. When a quantity of gas in our simulations begins non--linear collapse, the gas particles are replaced with a point mass once their density exceeds an arbitrary threshold. \cite{1994MNRAS.268..291W} make predictions about the radius the shocked shell should reach before it fragments, and the time after ignition of the radiation source at which this should occur. We therefore compare the radius and time at which the first point mass forms in each of our calculations with the predictions of W94 in Figures \ref{fig:rfrag} and \ref{fig:tfrag}.\\
\indent The fragmentation radii observed all lie inside the radius predicted by W94 and within $\sim25\%$ of the expected value. The observed fragmentation radii exhibit a spread amongst themselves of $\sim10\%$. The fragmentation times observed are also all less than the theoretical prediction by approximately $20\%$, again with an internal scatter of $\sim10\%$, and appear to converge as the particle number increases. We conclude that our calculations are adequately converged and that they have sufficient resolution to make a further study of the fragmentation process feasible. The fact that the shells in these simulations appear to fragment at earlier times (and therefore smaller radii) than predicted by W94 is not surprising, since fragmentation is in reality not an instantaneous process. In addition, there is the possibility that small quantities of noise in the initial particle distribution are likely to cause some regions of the shocked shell to fragment at earlier times (and thus smaller radii) than is predicted theoretically. We investigate this issue in the next section.\\

\section{Fragmentation}
\begin{figure*}
\begin{center}
\includegraphics[width=0.70\textwidth]{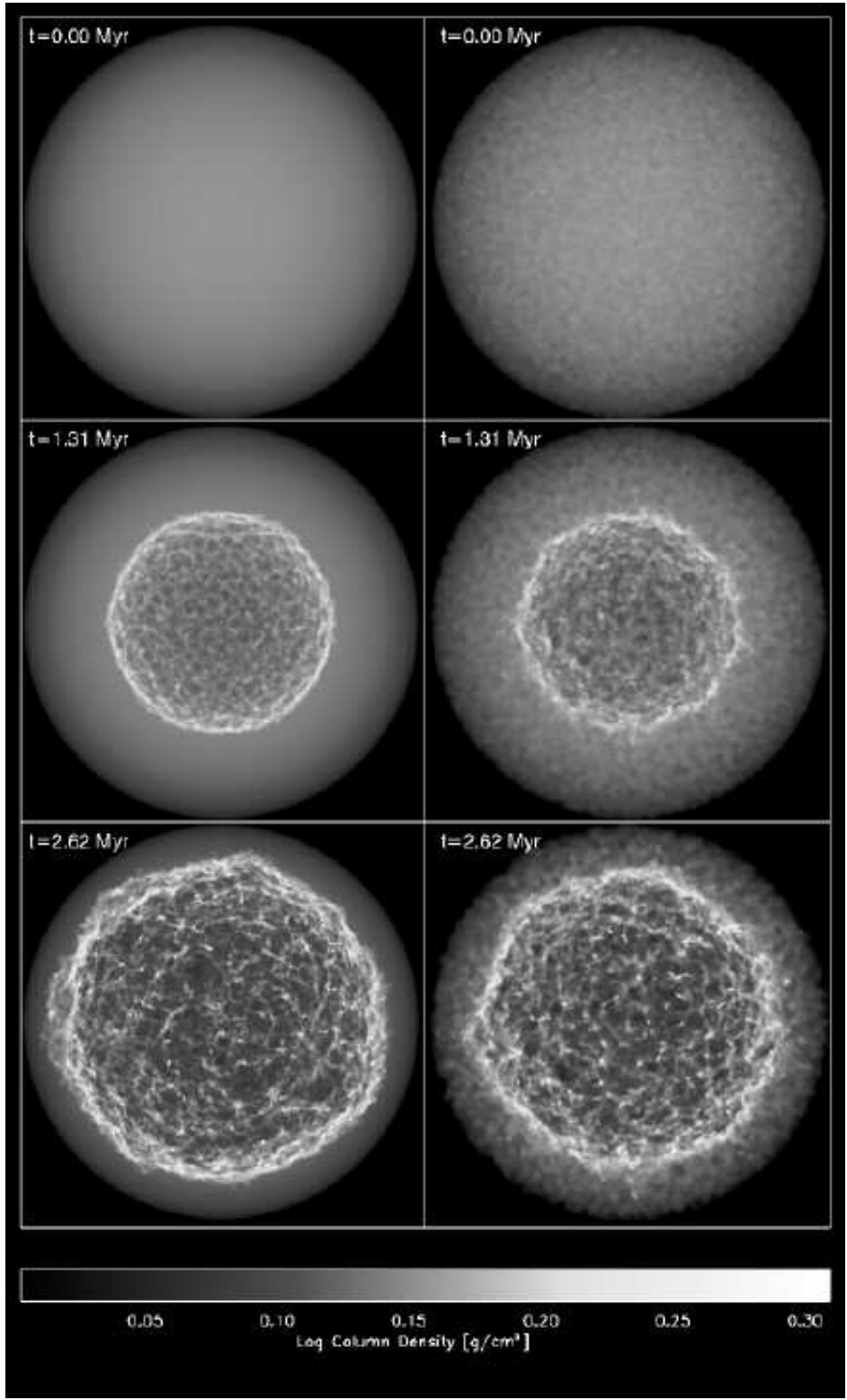}
\caption{Comparison of column--density maps (viewed along the z-axis) from three epochs of the `smooth' (left column) and `noisy' (right column) $1.3$ million particle calculations. Each image is $4.5\times10^{19}$ cm on a side. The noise introduced by randomising the initial particle positions is clearly visible in the images in the right column, although the morphology of the shocked shell is very similar in both calculations. To a large extent, the shock smoothes out the noise as it expands.}
\label{fig:random_compare}
\end{center}
\end{figure*}

The particles in all the simulations described above were initially distributed on uniform hexagonal--close--packed grids. To eliminate the possibility that the initial particle distribution has some effect on the fragmentation process, we repeated the $1.3\times10^{6}$ particle simulation with the particles distributed randomly. This produces fluctuations in gas density of a factor of $\sim2$. We compare screenshots from these two simulations in Figure \ref{fig:random_compare}. We find that the fragmentation time and radius for this calculation (determined in the same manner as above) are respectively $2.85\times10^{6}$ yr and $3.29\times10^{19}$ cm, very similar values to those obtained in the smooth simulation ($2.65\times10^{6}$ yr and $3.40\times10^{19}$ cm respectively). We conclude from these results that our results are robust against the presence of noise in the gas and that the initial particle distribution has no significant effect on the evolution of the system.\\ 

\begin{figure}
\includegraphics[width=0.5\textwidth]{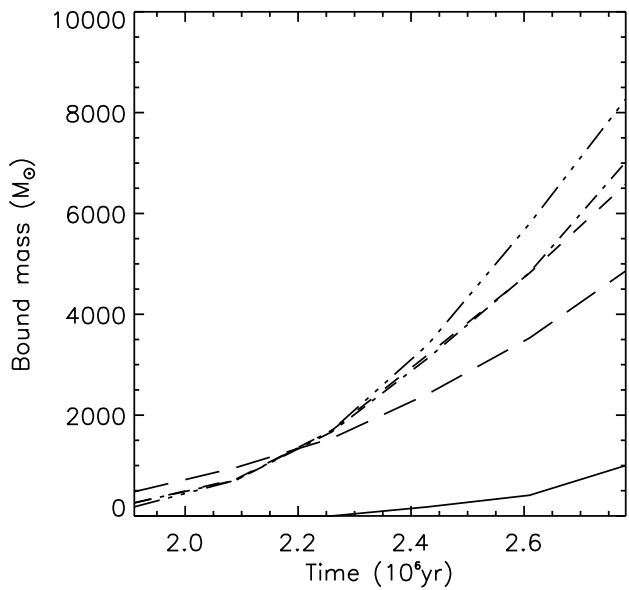}
\caption{Evolution with time of the quantities of self--gravitating material in all simulations (particle number expressed in units of $10^{6}$ particles): $0.37$ (solid line), $0.74$ (long--dashed line), $1.3$ (smooth) (short--dashed line), $1.3$ (noisy) (dot--dashed line) and $2.6$ (dot--dot--dot--dashed line).}
\label{fig:bound_mass}
\end{figure}

\begin{figure}
\includegraphics[width=0.5\textwidth]{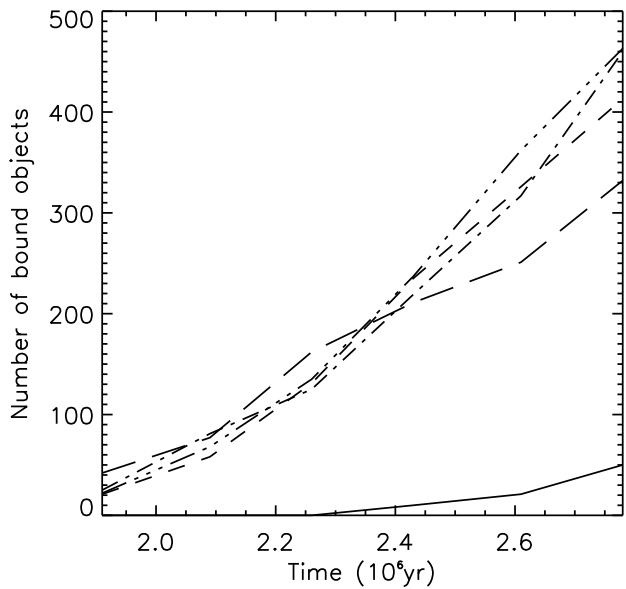}
\caption{Evolution with time of the number of self--gravitating clumps in all simulations (particle number expressed in units of $10^{6}$ particles): $0.37$ (solid line), $0.74$ (long--dashed line), $1.3$ (smooth) (short--dashed line), $1.3$ (noisy) (dot--dashed line) and $2.6$ (dot--dot--dot--dashed line).}
\label{fig:nclumps}
\end{figure}

\begin{figure*}
\begin{center}
\includegraphics[width=\textwidth]{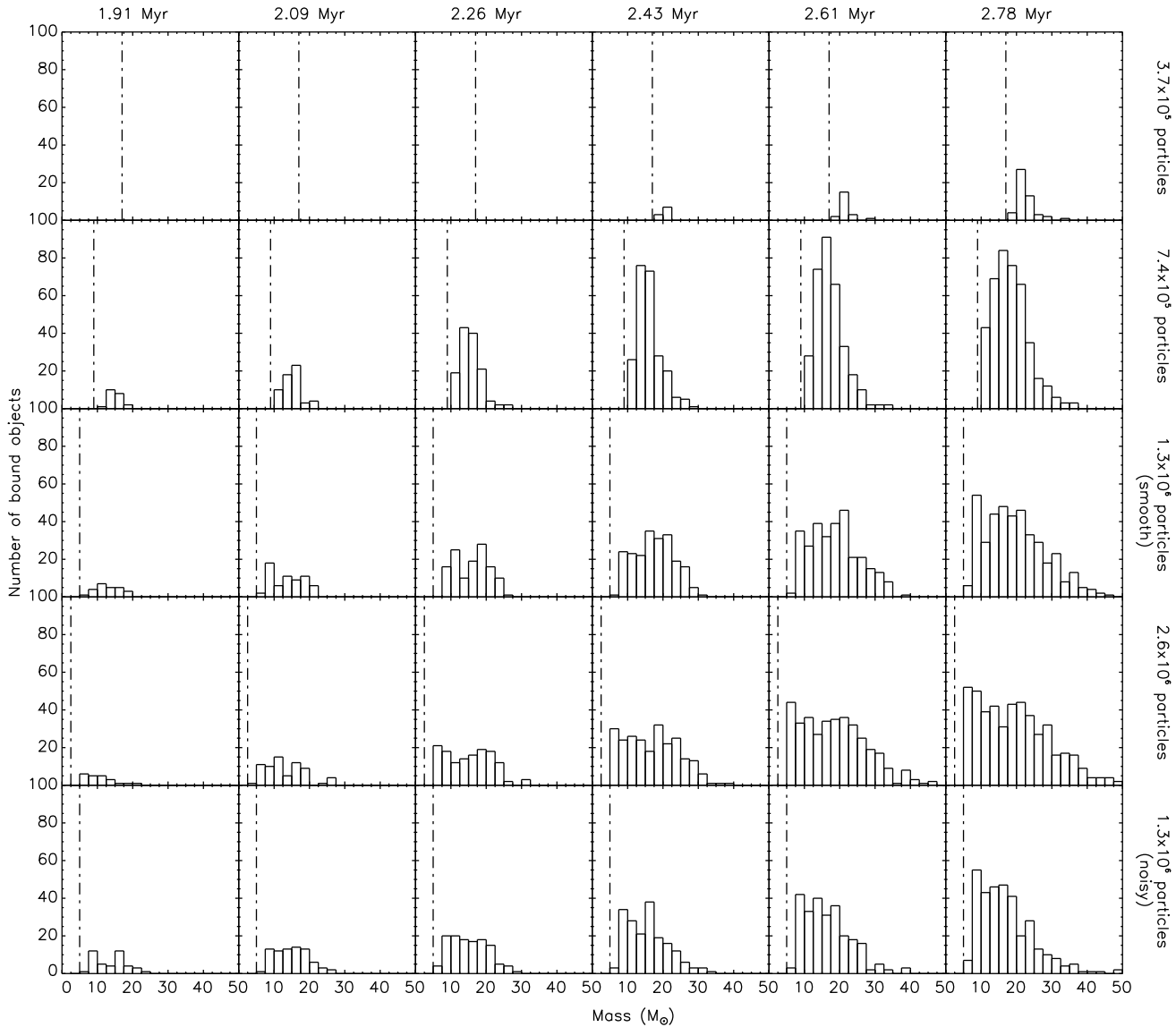}
\caption{The mass functions of bound clumps identified by the method described in the text at six epochs in all simulations. The epoch at which each mass function was generated is indicated across the top of the diagram, and the run to which it belongs is given on the right.}
\label{fig:mass_spectrum}
\end{center}
\end{figure*}

\indent Figure \ref{fig:random_compare} shows that the geometry of the shocked shell becomes extremely complex even before the first gravitationally--bound fragments form at $\sim2.7\times10^{6}$ yr. To study the process of fragmentation in detail, we devised a procedure to identify gravitationally--bound objects within our simulations. We first sorted all the SPH particles in order of decreasing density. We estimated the size of fragments approximately by eye from screenshots of our simulations and defined a cutoff radius $r_{cut}$. We experimented with values of $r_{cut}$ of $0.5$, $1.0$ and $2.0\times10^{18}$cm, finding that the choice had little influence on our results. We then descended through the sorted list of particles, taking each dense particle to be a possible seed for a bound clump. We fetched all the particles within $r_{cut}$ of each potential seed and sorted these by radius from the seed. We then added these particles to the seed one at a time and calculated the thermal, kinetic (in the centre--of--mass frame of the group of particles) and gravitational energies ($E_{therm}$, $E_{kin}$ and $E_{grav}$) of the collection of particles. This process was terminated when addition of the next particle would have rendered the group of particles unbound (i.e. when adding the particle would have have resulted in the ratio $(E_{therm}+E_{kin})/\left|E_{grav}\right|$ rising above a value of unity), thereby constructing bound objects around each potential seed. Particles assigned to clumps were flagged so that they could not be used as seeds themselves, and the sorted list was descended to an arbitrarily--defined lower threshold which was progressively lowered until convergence on the number and total mass of bound objects was obtained for each given snapshot from each simulation. In keeping with the mass resolution limit of \cite{1997MNRAS.288.1060B} mentioned earlier, clumps consisting of $<100$ particles were rejected.\\
\indent In Figure \ref{fig:bound_mass} we plot the increase with time of the total quantity of self--gravitating gas in all simulations (including the noisy $1.3\times10^{6}$--particle calculation), and in the $2.6\times10^{6}$--particle simulation for comparison. After a slow start, the amount of bound material in all three simulations rises approximately linearly with time. The two $1.3\times10^{6}$--particle calculations and the $2.6\times10^{6}$--particle calculation agree very well with each other, but the $0.74$ and in particular the $0.37$ million--particle calculations both show significantly less bound mass towards the ends of the simulations (by factors of about two and seven respectively). We observe that, in all cases, the amount of bound mass is only a fraction of the total mass in the shocked shell. Since fragmentation in both these simulations occurs at a radius of $\sim3.5\times10^{19}$ cm, the quantity of mass in the shell when fragmentation begins is $\sim3\times10^{4}$ M$_{\odot}$ and this mass clearly grows with time. We see therefore that only $\sim7\times10^{3}$ M$_{\odot}\sim10\%$ of the material in the shell is self--gravitating at the onset of fragmentation (a time of $\sim2.8\times10^{6}$ yr). The density of the shell at the onset of fragmentation is $\rho_{frag}\sim10^{-21}$ g cm$^{-3}$. The freefall time within the shell is then $(3\pi/(32G\rho_{frag}))^{(1/2)}\sim2.1\times10^{6}$ yr. Extrapolation of Figure \ref{fig:bound_mass} suggests that the quantity of bound mass will reach $\sim3\times10^{4}$ M$_{\odot}$ $\sim2\times10^{6}$yr after the onset of fragmentation, consistent with the freefall time in the shell. Due to the fact that the HII region reaches the edge of the simulation volume only $\sim0.2\times10^{6}$ yr after the onset of fragmentation, we cannot follow the fragmentation process to completion in these calculations.\\
\indent In Figure \ref{fig:nclumps}, we plot the increase with time of the number of self--gravitating clumps (identified by the method explained above) in all simulations. Given that \cite{1994MNRAS.268..291W} estimate the clump mass in these simulations to be $\sim50$ M$_{\odot}$, we should observe at least $\sim600$ clumps. We see that that the number of clumps rises approximately linearly with time (except in the lowest resolution run, which never forms enough clumps meeting our resolution limit to draw a conclusion of this nature). The two $1.3\times10^{6}$--particle calculations and the $2.6\times10^{6}$--particle calculation are again in very close agreement with each other and approach the numbers of clumps predicted by W94 at the ends of the calculations. The $0.74$ and particularly the $0.37\times10^{6}$--particle simulations produce fewer bound objects, by factors of less than two and about eight respectively, relative to the higher--resolution runs.\\
\indent In Figure \ref{fig:mass_spectrum} we plot the clump mass functions in all calculations at six epochs for comparison. The times at which the mass functions are taken are given across the top of the diagram and the calculation to which they belong is given on the right. The vertical dot--dashed lines represent the mass resolution limits of each calculation, taken to be the mass of $100$ particles. In common with Figures \ref{fig:bound_mass} and \ref{fig:nclumps}, the correspondence between the noisy and smooth $1.3$ million--particle runs and the $2.6$ million--particle run is very strong, implying both that we have achieved adequate numerical resolution and that noise in the initial conditions has little effect on the evolution of the system in question. The last two epochs of these runs in particular all show mass functions decreasing approximately linearly with mass, although the gradient is somewhat more negative in the calculation with primordial noise, resulting in slightly fewer fragments in the range $30-50$ M$_{\odot}$. By contrast, the mass functions generated in the $0.74$ million--particle run show more of a peaked morphology, with the peak remaining at a roughly constant mass of $\sim15$ M$_{\odot}$. Finally, the lowest--resolution calculation produces so few bound objects meeting the resolution criterion that the mass--functions generated can hardly be so called. This is a result of the relatively poor mass--resolution in this calculation. The $1.3$ and $2.6$ million--particle runs suggest that the majority of fragments formed in the system under investigation (within the timeframe of these simulations) have masses $<20$ M$_{\odot}$ and the mass resolution in the lowest--resolution run is $\sim17$ M$_{\odot}$, so it is not surprising that few fragments are observed in the $0.37$ million--particle calculation. The mean fragment mass observed in the $1.3$ and $2,6$ million--particle runs is $\sim20$ M$_{\odot}$, less than half the value of $\sim48$ M$_{\odot}$ predicted by W94, although these runs do produce a few objects of this mass. The density in the shell at the point of fragmentation, $\sim10^{-21}$ gcm$^{-3}$ implies a Jeans mass of $\sim40$ M$_{\odot}$, which is consistent with the fragment masses we observe and with the estimate of W94 to within a factor of two.\\
\indent 

\section{Discussion}
The convergence of our simulations on consistent values of $t_{frag}$ and $R_{frag}$ implies that we can adequately reproduce the gross evolution of this system for our particular choice of gas density and source luminosity. In addition, the similar evolution of the fragmenting shell in our two highest--resolution calculations implies that we have sufficient resolution to study the fragmentation process itself. Comparison of the high--resolution calculations with a third calculation in which we attempted to influence fragmentation by introducing random noise into the gas distribution shows that our results are robust against numerical noise and do not depend on the initial placement of the SPH particles. We conclude that we can usefully compare our results to the predictions of \cite{1994MNRAS.268..291W}.\\
\indent Our simulations yield values for $t_{frag}$, $R_{frag}$ which are respectively within $\sim20\%$ and $\sim25\%$ of those derived by W94. This implies that the criteria used by W94 to establish when and where the shell fragments are approximately correct. The fragment masses we observed are a factor of $\sim2.5$ smaller than predicted by W94. This discrepancy is not large, particularly when the necessarily approximate nature of their analysis is taken into account. However, it was not obvious at the outset that fragmentation in our simulations should proceed along the lines envisaged by W94. It may have been that the ionisation--front instabilities discussed by \cite{1979ApJ...233..280G} or noise in the initial gas distribution had some unforeseen effect on the fragmentation of the shell, but neither of these appear to be the case. Our simulations and the work of W94 imply that we have a good theoretical understanding of this (admittedly simple) problem and that we can move on to the study of more complex and realistic systems.\\
\indent The most obvious simplification made in this study is that the gas is initially uniform and smooth (although we showed that small amounts of noise in the gas do not affect the results substantially). Several authors have modelled the expansion of HII regions in gas distributions with power--law radial density profiles (e.g. \cite{1986MNRAS.221..635T}, \cite{1990ApJ...349..126F}) relevant to, for example, the expulsion of residual gas from proto--globular clusters. This work has not been performed in three--dimensions and the gravitational fragmentation of the gas has not been included. The work presented here implies that both these effects can now be included in realistic simulations.\\
\indent We have not explored the possibility that the existence of well--defined pre--existing clumps in the neutral gas may affect the fragmentation process. The passage of the ionisation front through or past such clumps may cause them to collapse, particularly if they are close to gravitational instability already. This is a different and more complex problem from the one we study here. The many possible characteristics of the clumps -- e.g. size, mass, density -- increase the size of the parameter space enormously. We therefore leave the study of this problem to a subsequent paper.\\
\indent We also neglect here the possibility that the stars formed by the collapsing fragments will themselves become ionising sources. The fragment mass estimated by Whitworth et al of $\sim48$ M$_{\odot}$ would produce an O--star with an ionising luminosity in excess of that of our source and, if we could follow the fragmentation process to completion, we may obtain hundreds of such objects.\\
\section{Conclusions}
We have shown that the analysis presented by W94 captures very well the evolution of a shell of self--gravitating gas surrounding an expanding HII region. This work and that of W94 shows that, from a theoretical standpoint, the collect--and--collapse model is a viable model for triggered star formation, at least when the gas from which the stars are forming is initially roughly uniform. Although the assumption that the gas is uniform is rather restrictive, the results of \cite{2005A&A...433..565D} suggest that the collect--and--collapse process does occur. They observed a total of seventeen HII regions. For two in particular, Sh 104 and RCW 79, they find a central ionising star with a roughly spherical HII region surrounded by a ring of dense molecular gas. In the case of RCW 79, there are clear signs that star formation has occurred and continues to occur in the molecular gas. They caution, however, that there are no signs of star formation in progress \textit{inside} the ring. This would be expected if the geometry were truly spherical -- Figure \ref{fig:random_compare} shows numerous dense fragments whose projected position is inside the shocked shell. This may suggest that the geometry of these objects is planar and that we observe them perpendicular to the plane. This conclusion was also reached by \cite{2000A&A...357.1001L} in their study of $\lambda$--Orionis. These authors further note that models in which the molecular material around $\lambda$--Ori was initially uniform do not well explain the region's present--day morphology. Such a system would make an interesting subject for further study.\\
\indent Regardless of this consideration, we can conclude that feedback from O--type stars can cause the fragmentation of even initially smooth and uniform--distributed gas. This is induced star formation in the strongest possible sense, since a smooth gas distribution will not otherwise fragment and form stars. This is an important conclusion, since it implies that any cloud of molecular gas can in principle be made to form one or more stars under the influence a nearby O--star.

\section{Acknowledgements}
JED acknowledges support from the University of Leicester's PPARC rolling grant. We would like to thank the referee for helpful comments.

\bibliography{myrefs}

\label{lastpage}

\end{document}